\documentclass{emulateapj}

\newcommand{\xte}{{\it RXTE}}

\newcommand{\epcs}{{\rm ergs\,cm^{-2}\,s^{-1}}}

\newcommand{\src}{IGR~J00291+5934}
\newcommand{\dist}{4}	

\newcommand{\stopdate}{6}	
\newcommand{\stopmjd}{53345}
\newcommand{\pubdate}{16}	
\newcommand{\pubmjd}{53355}

\slugcomment{Accepted by ApJL}

\shorttitle{Discovery of the millisecond X-ray pulsar IGR J00291+5934}
\shortauthors{Galloway et al.}

\begin{document}

\title{Discovery of the accretion-powered millisecond X-ray pulsar
IGR~J00291+5934}

\author{Duncan K. Galloway\altaffilmark{1},
   Craig B. Markwardt\altaffilmark{2,3},
   Edward H. Morgan\altaffilmark{1},
   Deepto Chakrabarty\altaffilmark{1,4},
and
   Tod E. Strohmayer\altaffilmark{2}}

\altaffiltext{1}{ Center for Space Research,
   Massachusetts Institute of Technology, Cambridge, MA 02139}

\email{duncan@space.mit.edu}

\altaffiltext{2}{ Laboratory for High Energy Astrophysics, NASA Goddard
Space Flight Center, Greenbelt, MD 20771}
\altaffiltext{3}{ Department of Astronomy, University of Maryland, College
Park MD 20742}
\altaffiltext{4}{ Department of Physics, Massachusetts Institute of
Technology, Cambridge, MA 02139}

\begin{abstract}
We report on 
observations of the sixth accretion-powered
millisecond pulsar, \src, with the {\em Rossi X-Ray Timing
Explorer}.  The source is a faint, recurrent X-ray transient initially
identified by {\em INTEGRAL}.  The 599 Hz (1.67 ms) pulsation had a
fractional rms amplitude of 8\% in the 
2--20~keV range, and its shape was approximately sinusoidal.
The pulses show an energy-dependent phase delay, with the 6--9~keV
pulses arriving up to 85~$\mu$s earlier than those at 
lower
energies.  No X-ray bursts, dips, or eclipses were detected.
The neutron star is in a circular 2.46~hr orbit with a very low-mass
donor, most likely a brown dwarf.  The binary parameters of the system
are similar to those of the first known accreting millisecond pulsar,
SAX J1808.4$-$3658. 
Assuming that the mass transfer is driven by gravitational radiation and
that the 2004 outburst fluence is typical, the 3-yr recurrence time implies a
distance of at least 
4~kpc.
\end{abstract}


\section{Introduction}

The growing sample of accretion-powered millisecond X-ray pulsars 
divides naturally into two groups. Three of the five sources known to date
(XTE~J1751$-$305, \citealt{markwardt02}; XTE~J0929$-$314,
\citealt{gal02d}; and XTE~J1807$-$294, \citealt{markwardt03}) are in
ultracompact binaries with orbital periods of $\approx40$~min. The Roche lobes
in such tiny binaries cannot contain a main-sequence companion, indicating
that the mass donors are highly evolved and 
H-poor 
\cite[]{nrj86,nr03,db03}. The other two (SAX~J1808.4$-$3658,
\citealt{wij98b,chak98d};
and XTE~J1814$-$338, \citealt{markwardt03a}) have orbital periods of 2.01
and 4.28~hr, respectively, and have H-rich donors
\cite[likely a brown dwarf in the case of
SAX~J1808.4$-$3658;][]{bc01}.  The latter two sources have also both
exhibited thermonuclear (type-I) X-ray bursts,
whereas the three ultracompact
binaries have not. 
All five pulsars are soft X-ray transients, with outburst durations of
order weeks and recurrence times of order years.
It remains unclear why millisecond pulsations are easily detectable in
these five sources, but not in the over 50 other neutron stars in low-mass
X-ray binaries 
\cite[e.g.][]{czb01,tcw02}.

The 2002 October launch of the {\it INTEGRAL}\/ gamma-ray mission
\cite[]{integral03}, with its wide field of view and good sensitivity to
hard X-ray sources, has provided a new avenue for detections of X-ray
transients.
{\it INTEGRAL}\/ discovered the new X-ray transient \src\ ($l=120\fdg1$,
$b=-3\fdg2$) 
on 2004 Dec. 2 \cite[MJD 53341;][]{eckert04}.  Followup {\it Rossi
X-ray Timing Explorer}\/ ({\it RXTE}) observations revealed pulsations
at a frequency of 598.88~Hz \cite[]{mark04a}, and
exhibiting a sinusoidal frequency modulation indicative of a 147.4~min
orbit \cite[]{mark04b}.
The detection of pulsations motivated extensive multiwavelength followup.
An $R\approx17.4$ candidate optical counterpart was identified within the
{\it INTEGRAL}\/ error circle, at
R.A. = $00^{\mathrm h}29^{\mathrm m}03\fs06$,
decl. = $+59\arcdeg34\arcmin19\farcs0$
\cite[J2000.0, $1\sigma$ undertainty radius $0\farcs25$;][]{fox04}.
Optical spectroscopy of the candidate revealed weak He{\sc ii} and
H$\alpha$ emission, supporting its association with \src\
\cite[]{roelofs04}.  Radio observations also revealed evidence for
variable emission consistent with the counterpart position
\cite[]{pooley04,fender04}.

In this letter we describe the detailed analysis of the \xte\/ observations 
following the discovery.

\section{Observations}

We obtained a
series of pointed {\em RXTE}\/ observations of \src\ between 2004
December 3--\pubdate\ (MJD 53342--\pubmjd),
with a total exposure of approximately
295~ks. Our analysis is primarily
based on data from the {\em RXTE}\/ Proportional Counter Array
\cite[PCA][]{xte96}, which consists of five identical gas-filled
proportional counter units (PCUs) sensitive to X-ray photons in the
2.5--60~keV range, with a total effective area of $\approx6000\ {\rm
cm}^2$.  The data were collected either in GoodXenon or generic Event
modes (in addition to the standard data modes). GoodXenon records the
arrival time (1~$\mu$s resolution) and energy (256 channel resolution) of
every unrejected photon, while the chosen Event configuration has $125\
\mu$s time resolution and 64 energy channels. We analysed FITS production
data using {\sc LHEASOFT} version 5.3.1 (2004 May 20) and estimated the
background flux using the ``CM'' bright-source model for PCA gain epoch 5
(from 2000 May 13).  The photon arrival times at the spacecraft were
converted 
 \centerline{\epsfxsize=8.5cm\epsfbox{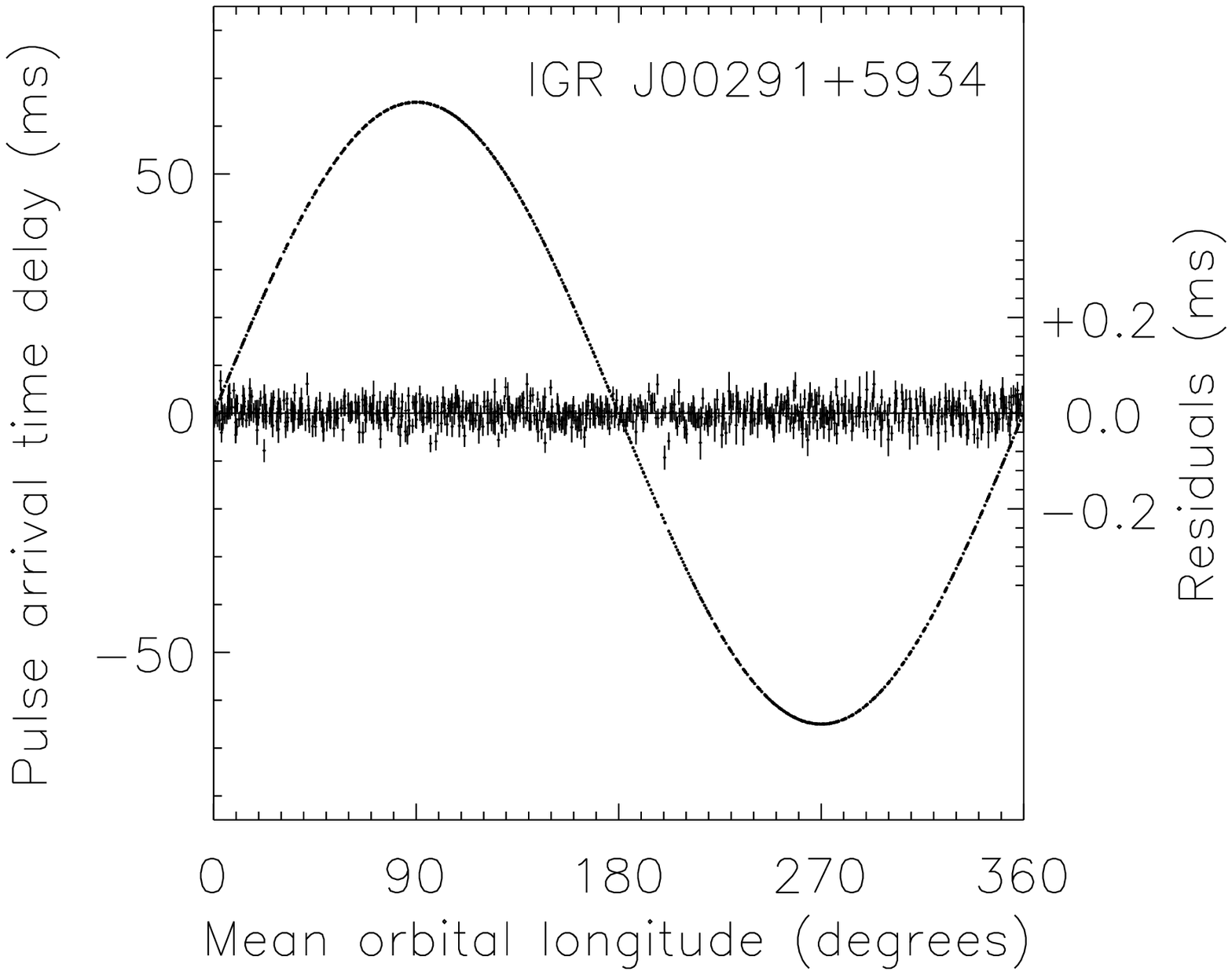}}
 \figcaption[]{Pulse arrival time delays in \src\ due to the 2.46~hr
binary orbit, with respect to a constant pulse frequency model. 
The measured time delays are plotted as filled circles, and the symbols
with error bars (representing the $1\sigma$ uncertainties) show the model
residuals (on a $100\times$ expanded scale, right-hand $y$-axis).
 \label{tdelays} }
\noindent to barycentric dynamical times (TDB) at the solar system
barycenter using the Jet Propulsion Laboratory DE-200 solar system
ephemeris \cite[]{standish92} along with a spacecraft ephemeris and {\em
RXTE}\/ fine clock corrections.  The position adopted was that of the
proposed optical counterpart \cite[]{fox04}.  Data from the 20--200~keV
High-Energy X-ray Timing Experiment \cite[HEXTE;][]{hexte96} were also
used to characterize the hard X-ray spectrum.

\section{Analysis and results}
\label{sec1}

For our timing analysis, we selected photons from the top layer of each
PCU in the energy range 3--13~keV (absolute channels 7--30) in order
to maximize the pulsed signal to noise ratio.
We binned the arrival times for these photons into $2^{-13}$~s
($\approx$0.12~ms) samples. A 599~Hz pulsed signal was easily detectable
in these data, with a fractional rms amplitude of between 7--9\%. 
The rms amplitude decreased throughout the outburst, gradually to 6\% on
December 11 (MJD~53350) and then more rapidly down to a minimum of
$\approx2$\% by December 14.
The pulsations became undetectable ($<1$\% rms) from December 15
(MJD~53354) onwards, and the residual flux which was detected after this
date likely arose from other sources in the \xte\/ field of view.
The pulse frequency, measured every 64~s, exhibited a sinusoidal
modulation attributed to Doppler shifts from the binary orbit. We
fitted a preliminary orbital model to the frequency measurements between
December 3--\stopdate\ (MJD 53342--\stopmjd),
and then used this preliminary model in a more precise pulse phase
analysis 
\cite[e.g.][]{mt77}.  
We first corrected the time series
using the preliminary orbit. 
We then folded 64-s intervals of data on the pulsar period estimated from
the frequency fit, and determined the epoch of maximum flux within each
interval.  We fitted the residuals (measured pulse arrival time minus the
predicted arrival time) to compute differential corrections to our
Keplerian orbit model \cite[see, e.g.][]{dpb81}, and repeated the process
with the corrected model parameters until no further reduction in the
rms residual error was achieved.
No significant eccentricity was detectable.  Our best-fit orbit and spin
parameters for the pulsar are given in Table~1.  The pulse time delays due
to the orbit are shown in Fig.~\ref{tdelays}.

We extracted light curves in 12 energy bands between 2--20~keV with
0.12~ms time bins
from the observations between December 3--\stopdate\ (MJD~53342--\stopmjd).
We then corrected these light curves for orbital motion based on the
best-fit orbital solution, and folded on the pulsar period to measure the
energy dependence of the pulsations.
The fractional rms pulsed amplitude
decreased with energy, from 10\% in the 2--3~keV range to
6.5\% between 7--16~keV, dropping further to 5\% above 16~keV (Fig.
\ref{prof}, upper panel).  
The profile was approximately sinusoidal at all energies. Of the higher
harmonics, only the
second harmonic (1198~Hz) was detected at $>3\sigma$ confidence in more
than one energy band (although not in all bands), at 
fractional rms amplitudes
of a few tenths of a percent.
Upper limits on the fractional amplitudes for the third and higher
harmonics were of similar magnitude.
The pulse phase had an energy dependence.  The hard
photons arrived earlier than the soft, with a maximum lead of 
around 90~$\mu$s (equivalent to 5\% of the pulse period) at
6--9~keV (Fig. \ref{prof}, lower panel).
In contrast to other accretion-powered millisecond pulsars, where the hard
lead saturates above 10~keV \cite[][]{cui98,gal02d}, in \src\ we found
evidence for a slight decrease in the lead above 9~keV.
We found no evidence for modulation of the X-ray intensity at the orbital
period, ruling out eclipses or dipping behavior.
We searched for kilohertz quasi-periodic oscillations (kHz QPOs) in the range
50--4000~Hz, averaging FFTs from 128 and 512-s segments of data over
entire observations and entire days, but found no significant peaks.
We estimate a typical $3\sigma$ upper limit for features with FWHM of up
to 100~Hz of $\la1$\% rms \cite[compare to the 5--10\%~rms kHz QPOs
measured in SAX~J1808.4$-$3658;][]{wij03}.
We also searched each observation of \src\ for thermonuclear X-ray bursts,
but found none\footnote{
A burst was detected by
both PCA and HEXTE at 18:34:21~UT during an observation of \src\ on 2004
December 12. However, this event had a very unusual energy spectrum and
was probably a bright gamma-ray burst 
observed through the side of the \xte\/ detectors.
The burst was also detected by {\em INTEGRAL} (SPI/ACS monitor trigger
4779) and the {\em Mars Odyssey} mission
\cite[]{hurley02}, and the
derived interplanetary network (IPN3; see
{\url http://www.ssl.berkeley.edu/ipn3/})
position
annulus excludes \src\ as the origin (K. Hurley, pers. comm.).
We therefore designate this event as GRB041212.
}.

We fitted HEXTE and PCA spectra, the latter extracted separately for
each PCU (excluding PCU \#0, which has lost its propane veto layer)
between 2.5--200~keV. We obtained a good fit with a model consisting of a
power law with spectral index 1.7--1.8, attenuated by neutral absorption with a
column depth of $\approx10^{22}\ {\rm cm^{-2}}$. This is only slightly
higher than the survey value of $4.7\times10^{21}\ {\rm cm^{-2}}$
\cite[]{dl90}.
A Comptonization model \cite[]{tit94} gave a marginally better fit to the
highest signal-to-noise spectra
but the scattering electron temperature (which gives rise to a high-energy
roll off in the spectra) could not be measured, save that it was
$\ga60$~keV.
This implies an optical depth for scattering (assuming spherical geometry)
of $\tau=0.8$.
We also found evidence for neutral Fe K$\alpha$ line 
 \centerline{\epsfxsize=8.5cm\epsfbox{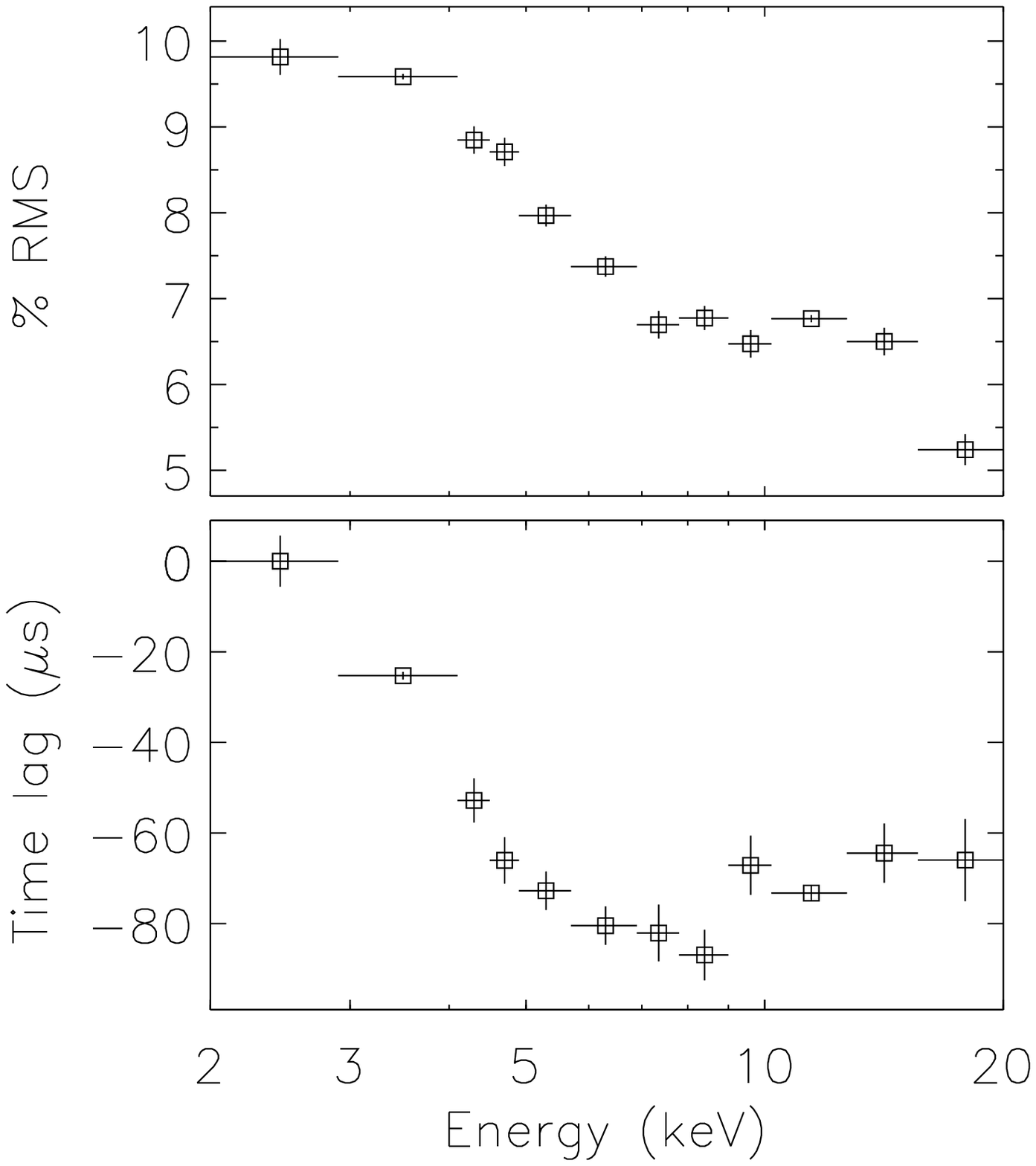}}
  \figcaption[]{{\it Top panel}\/ Fractional rms amplitude of pulsations in
\src\ as a function of photon energy. {\it Bottom panel}\/ Pulse arrival
time as a function of energy, measured relative to the pulsation in the
2--3~keV energy band (negative values correspond to earlier arrival
times).  In both panels, vertical error bars indicate the $1\sigma$
uncertainties
 \label{prof} }
\vspace{0.1cm}
\noindent emission at 6.4~keV,
with equivalent width $\approx50$~eV.
While the line emission could potentially originate from other sources
within the $1^\circ$  \xte\/ field of view, or from diffuse Galactic
emission, a {\it Chandra}\/ observation on December 14 also found evidence
for Fe line emission in the zeroth order spectrum (M. Nowak, pers. comm.).
In order to minimise the fit residuals for the PCA data we also added two
edge features, one at 7.12~keV (Fe neutral K-edge) and one at 4.7~keV
(instrumental xenon edge).

An X-ray intensity history of the 2004 outburst is available both from the
PCA observations and from the \xte\/ All Sky Monitor \cite[ASM;][]{asm96};
these are shown in Fig. \ref{history}.
The flux measured by the PCA decreased from $1.14\times10^{-9}\ \epcs$ to 
$\la10^{-10}\ \epcs$ (2.5--25~keV) between 2004 December
3--15 (MJD~53342--53354).
During the discovery observations by {\it INTEGRAL}\/ on December 2
(MJD~53341) the JEM-X instrument measured a 3--10~keV flux of
$23\pm5$~mCrab, while the ISGRI instrument measured $55\pm5$~mCrab in the
20--60~keV band \cite[]{eckert04}.
For an absorbed power law with $n_{\rm H}=10^{22}\ {\rm cm^{-2}}$ and
photon index $\Gamma=1.7$ (where $dN/dE\propto E^{-\Gamma}$), this
translates to a 2.5--25~keV flux of 1.0--$1.1\times10^{-9}\ \epcs$. Thus,
the peak PCA flux measurement of
$(1.14\pm0.06)\times10^{-9}\ \epcs$
(2.5--25~keV) on December 3 (MJD~53342) is likely a good measure of the peak outburst flux.
We calculated a correction factor of 
2.54
from the broadband spectral fits
to apply to the 2.5--25~keV fluxes, in order to estimate the bolometric
flux (here taken as the flux from 0.1--200~keV).  Assuming the outburst
commenced on December 2 (MJD~53341),
we estimated a total fluence for
the outburst of 
$1.8\times10^{-3}\ {\rm ergs\,cm^{-2}}$.

The \xte\/ mission-long (1996--2004) 2--10~keV flux history of \src\ produced
from ASM observations confirmed the current outburst, as well as
revealing evidence for two previous instances of activity, on 1998
November 26--28 and 2001 September 11--21 \cite[]{remillard04}. These
results indicate that the transient appears regularly, roughly every three
years, similar to the 401~Hz pulsar SAX~J1808.4$-$3658, which exhibits
outbursts every $\approx2$~yr \cite[e.g.][]{wij03a}.
Combined with the estimated 3-yr recurrence time of the outbursts in \src, and
assuming similar fluences, we estimate a long-term time-averaged flux for
\src\ of
$1.8\times10^{-11}\ \epcs$.

\section{Discussion}
\label{sec3}

With a spin frequency of 599 Hz, \src\ is the fastest
known accretion-powered millisecond pulsar.  Although it is neither
the fastest known pulsar \cite[cf. radio pulsar PSR B1937+21, $\nu=641$
Hz;][]{backer82}
nor the fastest spinning neutron star in an X-ray
binary (cf. nuclear-powered millisecond pulsar 4U 1608$-$52, $\nu=619$ Hz;
Hartman et al. 2005, in prep.), it is part of a growing population of
pulsars found in the 600--650~Hz range.  These objects underscore 
the statistically significant absence of pulsars with spins faster than
650~Hz which has been noted previously \cite[]{chak03a,chak05a}.
It remains unclear whether this cutoff in the spin distribution
reflects the distribution of neutron star magnetic field strengths
\cite[e.g.][]{wz97}
or angular momentum losses from gravitational
radiation
\cite[e.g.][]{wagoner84,bil98c}

 \centerline{\epsfxsize=8.5cm\epsfbox{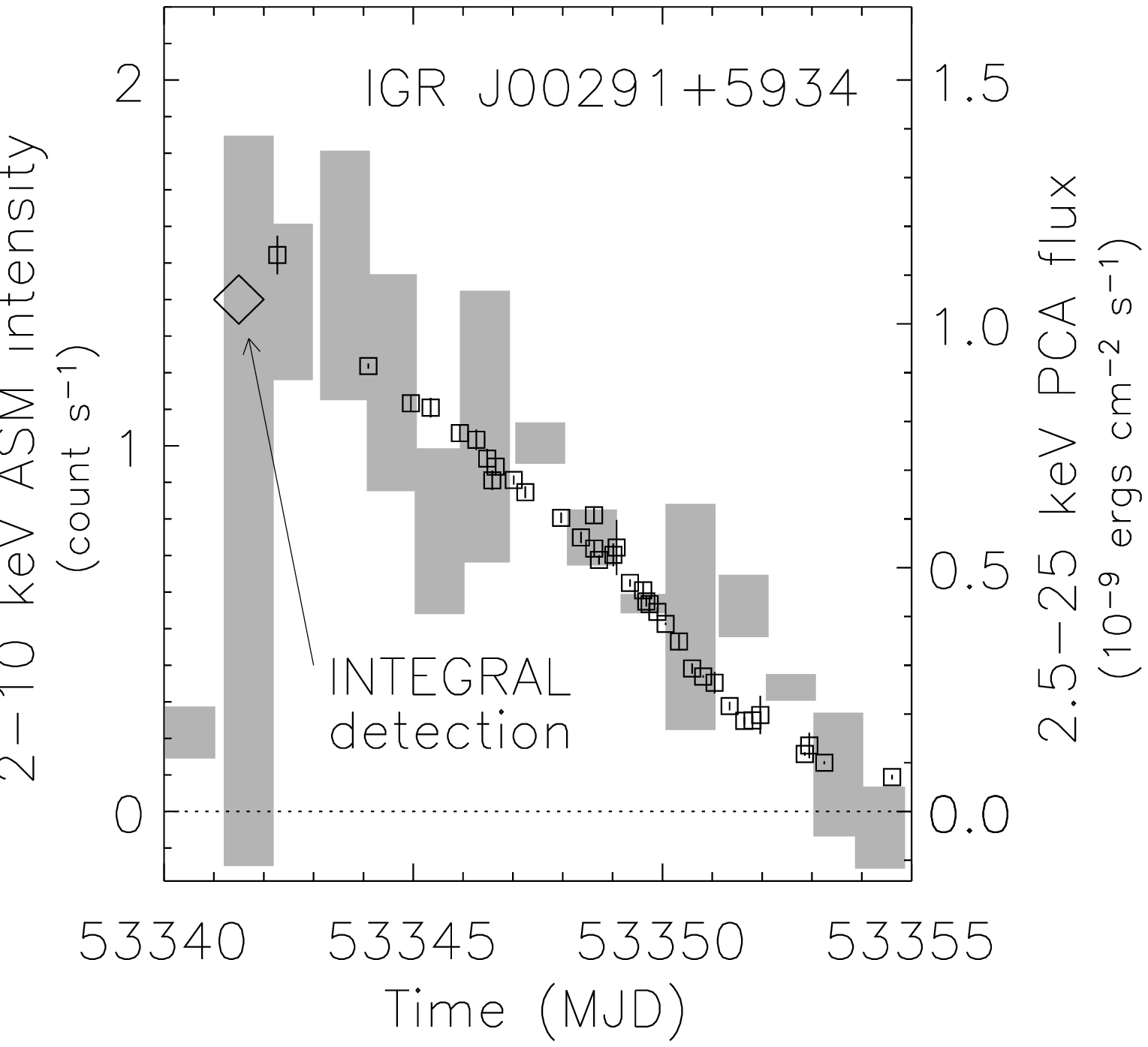}}
 \figcaption[]{X-ray intensity of \src\ throughout the 2004 outburst. The
1-day averaged ASM intensity values and their $1\sigma$ confidence
intervals are shown as shaded regions (left-hand $y$-axis) while the
integrated 2.5--25~keV PCA flux measurements are overplotted as open
squares (right-hand $y$-axis). The $1\sigma$ uncertainties on the PCA flux
measurements are shown, but are generally smaller than the symbols.
The estimated 2.5--25~keV flux during the intitial {\it INTEGRAL}\/
detection is indicated by the open diamond.
 \label{history} }
In many respects, \src\ closely resembles the best-known millisecond
pulsar, SAX~J1808.4$-$3658. The mass function in \src\ implies a minimum
companion mass of $M_c=0.039M_\odot$ (for a $1.4M_\odot$ neutron star and
binary inclination $i=90^\circ$, although the lack of X-ray eclipses or dips
implies $i\la85^\circ$). For an isotropic {\it a priori} distribution of
inclinations, we find an upper limit on the companion mass
of $0.16M_\odot$ (95\% confidence, assuming $M_x=2M_\odot$). The inferred
mass-radius relation intersects that of a hydrogen main sequence star at
a companion mass of $0.24M_\odot$ (Fig. \ref{ev}), which requires an
improbably small inclination of $10^\circ$. This
indicates that, like SAX~J1808.4$-$3658,
the mass donor in \src\ is a
hot brown dwarf, likely heated by low-level X-ray emission during 
quiescence \cite[]{bc01}.
For very low-mass companions,
the mass transfer rate driven by gravitational radiation is
\begin{equation}
\dot{M}_{\rm GR} \simeq 3.3\times10^{-12}
                     \left(\frac{M_c}{0.039M_\odot}\right)^2
           \left(\frac{M_x}{1.4M_\odot}\right)^{2/3}
  \left(\frac{P_{\rm orb}}{2.46\ {\rm hr}}\right)^{-8/3}\ M_\odot\,{\rm yr^{-1}}
\end{equation}
where $M_c$ is the mass of the donor and $M_x$ the mass of the neutron
star. Using the long-term averaged flux of 
$1.8\times10^{-11}\ \epcs$,
we derive a minimum distance of roughly \dist~kpc (for a companion mass of
$0.039M_\odot$). 
This would
place the system at least 
200~pc away from the Galactic plane, likely within the disk.
We note that the $\dot{M}_{\rm GR}$ above refers to the mass transfer
rate averaged over a much longer time scale than the \xte\/ mission
lifetime, so that this distance estimate must be viewed with some caution.

The peak inferred accretion rate for a distance of \dist~kpc is
$\ga0.04\,d_{\rm 4\,kpc}^2\ \dot{M}_{\rm Edd}$
(where $\dot{M}_{\rm Edd}=2\times10^{-8}\ M_\odot\,{\rm yr^{-1}}$ is
the hydrogen Eddington limit for spherical
accretion onto a $1.4M_\odot$
neutron star). This is 
comparable to
the maximum reached by SAX~J1808.4$-$3658 during its 2002 October
outburst, when four thermonuclear (type-I) X-ray bursts were detected by
\xte\ \cite[]{chak03a}. The minimum burst interval in SAX~J1808.4$-$3658
was 21~hr.
If the distance to \src\ is greater than \dist~kpc, the accretion rate
would also be larger by a factor of $(d/4\,{\rm kpc})^2$, and we would
expect more frequent bursts \cite[e.g.][]{ramesh03}.  In this case,
however, we would expect to have detected at least one burst in our
 \centerline{\epsfxsize=8.5cm\epsfbox{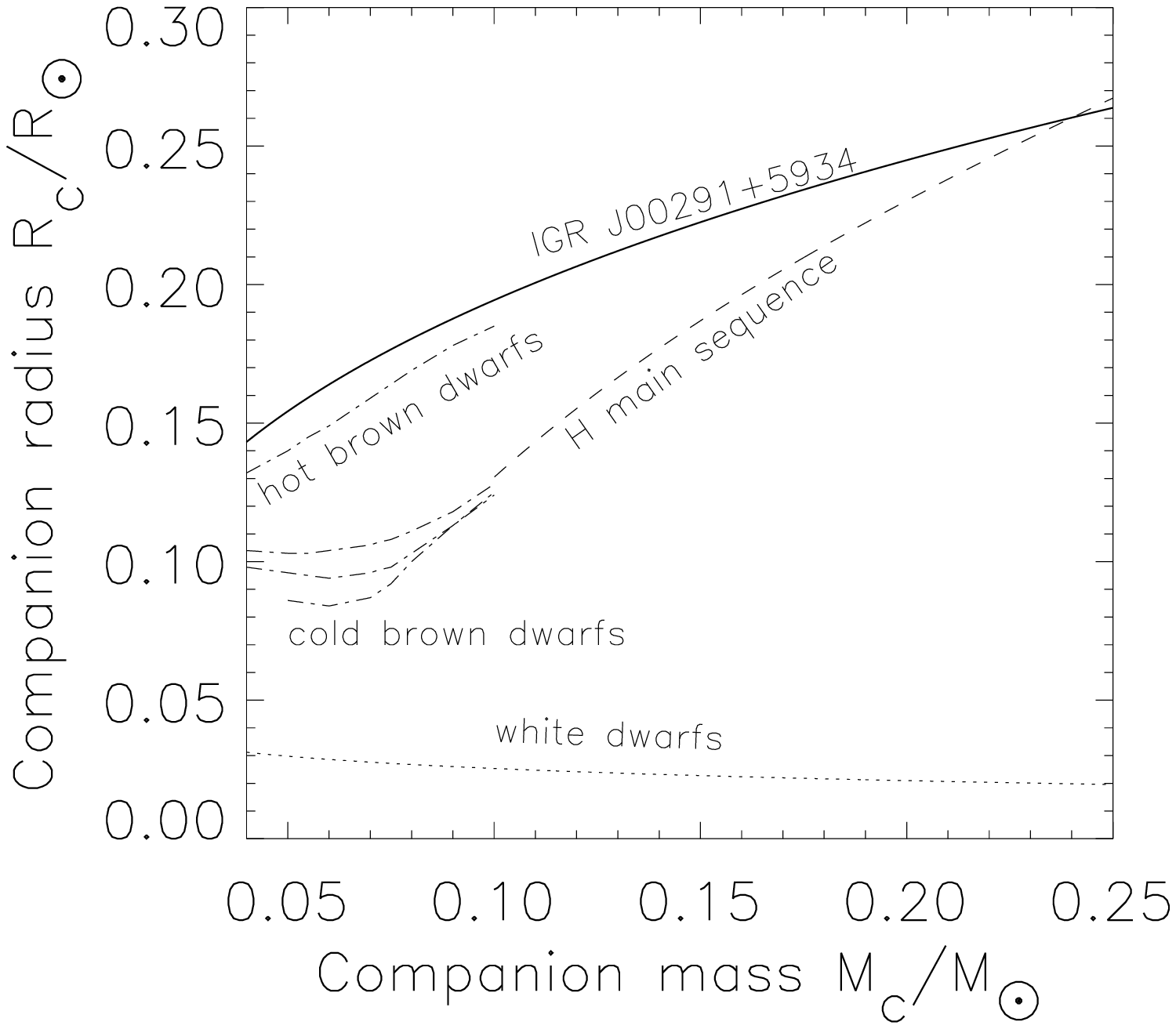}}
 \figcaption[]{Possible companion types for \src. The dark solid curve
shows the allowed mass-radius relation based on the \xte\/ observtions.
The other curves show theoretical mass-radius relations for cold He white
dwarfs ({\it dotted curve}), low-mass hydrogen main sequence stars ({\it
dashed curve}), and brown dwarfs ({\it dot-dashed curves}) of ages 0.1,
0.5, 1.0 and 5.0~Gyr from top to bottom. The young (0.1~Gyr) brown dwarf model
should be roughly representative
       of a bloated, X-ray-heated brown dwarf donor \cite[see][]{bc01}
 \label{ev} }
\noindent\xte\ observations, since the observational duty cycles during the 2002
October outburst of SAX~J1808.4$-$3658 
and the 2004 December outburst of
\src\ were comparable. The non-detection of thermonuclear bursts
from \src\ thus suggests that the source distance is not
significantly greater than \dist~kpc.

\acknowledgments

We are grateful to Jean Swank, Evan Smith, and the \xte\/ operations team
at NASA Goddard Space Flight Center for their help in scheduling these
observations so promptly. We also thank Lars Bildsten, Jacob Hartman, and
Alan Levine for useful discussions.
This research has made use of data obtained through the High Energy
Astrophysics Science Archive Research Center Online Service, provided by
the NASA/Goddard Space Flight Center.  This work was supported in part by
the NASA Long Term Space Astrophysics program under grant NAG 5-9184, and
the NASA \xte\/ Guest Observer Program.

\begin{deluxetable}{lr}
\tabletypesize{\small}
\tablecolumns{2}
\tablewidth{260pt}
\tablecaption{Orbit and Spin Parameters for \src}
\tablehead{ \colhead{Parameter} & \colhead{Value}}
\startdata
Orbital period, $P_{\rm orb}$ (s) & 8844.092(6) \\
Projected semimajor axis, $a_{\rm x}\sin i$ (light-ms)& 64.993(2) \\
Epoch of 90$^\circ$ mean longitude, $T_{\pi/2}$ (MJD/TDB) & 53345.1875164(4) \\
Spin frequency, $\nu_0$ (Hz) & 598.89213064(1) \\
Mass function $f_X$ ($M_\odot$) & $2.81311(7)\times10^{-5}$ \\
Eccentricity, $e$ & $<2\times10^{-4}$ (3$\sigma$) \\
Spin frequency derivative, $|\dot\nu|$ (Hz~s$^{-1}$)& $<8\times 10^{-13}$
(3$\sigma$)\\
\\
$\chi^2$/dof & 980/748 \\
\enddata
\tablecomments{These parameters assume the position of the optical
counterpart identified by \cite{fox04}. Our timing analysis is based on
the first 50~ks of data, between 2004 December 3--\stopdate\ (MJD 53342--\stopmjd).
}
\end{deluxetable}

\clearpage

\clearpage


\begin{thebibliography}{39}
\expandafter\ifx\csname natexlab\endcsname\relax\def\natexlab#1{#1}\fi

\bibitem[{{Backer} {et~al.}(1982){Backer}, {Kulkarni}, {Heiles}, {Davis}, \&
  {Goss}}]{backer82}
{Backer}, D.~C., {Kulkarni}, S.~R., {Heiles}, C., {Davis}, M.~M., \& {Goss},
  W.~M. 1982, \nat, 300, 615

\bibitem[{{Bildsten}(1998)}]{bil98c}
{Bildsten}, L. 1998, \apjl, 501, L89

\bibitem[{{Bildsten} \& {Chakrabarty}(2001)}]{bc01}
{Bildsten}, L. \& {Chakrabarty}, D. 2001, \apj, 557, 292

\bibitem[{{Chakrabarty}(2005)}]{chak05a}
{Chakrabarty}, D. 2005, in Binary Radio Pulsars, ed. F.~A. {Rasio} \& I.~H.
  {Stairs} (San Fransisco: ASP Conf. Ser. 328), 279 (astro--ph/0408004)

\bibitem[{{Chakrabarty} \& {Morgan}(1998)}]{chak98d}
{Chakrabarty}, D. \& {Morgan}, E.~H. 1998, \nat, 394, 346

\bibitem[{{Chakrabarty} {et~al.}(2003){Chakrabarty}, {Morgan}, {Muno},
  {Galloway}, {Wijnands}, {van der Klis}, \& {Markwardt}}]{chak03a}
{Chakrabarty}, D., {Morgan}, E.~H., {Muno}, M.~P., {Galloway}, D.~K.,
  {Wijnands}, R., {van der Klis}, M., \& {Markwardt}, C.~B. 2003, \nat, 424, 42

\bibitem[{{Cui} {et~al.}(1998){Cui}, {Morgan}, \& {Titarchuk}}]{cui98}
{Cui}, W., {Morgan}, E.~H., \& {Titarchuk}, L.~G. 1998, \apjl, 504, L27

\bibitem[{{Cumming} {et~al.}(2001){Cumming}, {Zweibel}, \& {Bildsten}}]{czb01}
{Cumming}, A., {Zweibel}, E., \& {Bildsten}, L. 2001, \apj, 557, 958

\bibitem[{{Deeter} {et~al.}(1981){Deeter}, {Pravdo}, \& {Boynton}}]{dpb81}
{Deeter}, J.~E., {Pravdo}, S.~H., \& {Boynton}, P.~E. 1981, \apj, 247, 1003

\bibitem[{{Deloye} \& {Bildsten}(2003)}]{db03}
{Deloye}, C.~J. \& {Bildsten}, L. 2003, \apj, 598, 1217

\bibitem[{{Dickey} \& {Lockman}(1990)}]{dl90}
{Dickey}, J.~M. \& {Lockman}, F.~J. 1990, \araa, 28, 215

\bibitem[{{Eckert} {et~al.}(2004){Eckert}, {Walter}, {Kretschmar}, {Mas-Hesse},
  {Palumbo}, {Roques}, {Ubertini}, \& {Winkler}}]{eckert04}
{Eckert}, D., {Walter}, R., {Kretschmar}, P., {Mas-Hesse}, M., {Palumbo}, G.
  G.~C., {Roques}, J.-P., {Ubertini}, P., \& {Winkler}, C. 2004, The
  Astronomer's Telegram, 352

\bibitem[{{Fender} {et~al.}(2004){Fender}, {de Bruyn}, {Pooley}, \&
  {Stappers}}]{fender04}
{Fender}, R., {de Bruyn}, G., {Pooley}, G., \& {Stappers}, B. 2004, The
  Astronomer's Telegram, 361

\bibitem[{{Fox} \& {Kulkarni}(2004)}]{fox04}
{Fox}, D.~B. \& {Kulkarni}, S.~R. 2004, The Astronomer's Telegram, 354

\bibitem[{{Galloway} {et~al.}(2002){Galloway}, {Chakrabarty}, {Morgan}, \&
  {Remillard}}]{gal02d}
{Galloway}, D.~K., {Chakrabarty}, D., {Morgan}, E.~H., \& {Remillard}, R.~A.
  2002, \apjl, 576, L137

\bibitem[{{Gruber} {et~al.}(1996){Gruber}, {Blanco}, {Heindl}, {Pelling},
  {Rothschild}, \& {Hink}}]{hexte96}
{Gruber}, D.~E., {Blanco}, P.~R., {Heindl}, W.~A., {Pelling}, M.~R.,
  {Rothschild}, R.~E., \& {Hink}, P.~L. 1996, \aaps, 120, C641

\bibitem[{{Hurley} {et~al.}(2002){Hurley}, {et~al.}}]{hurley02}
{Hurley}, K. {et~al.} 2002, GCN Circ. 1372
 
\bibitem[{{Jahoda} {et~al.}(1996){Jahoda}, {Swank}, {Giles}, {Stark},
  {Strohmayer}, {Zhang}, \& {Morgan}}]{xte96}
{Jahoda}, K., {Swank}, J.~H., {Giles}, A.~B., {Stark}, M.~J., {Strohmayer}, T.,
  {Zhang}, W., \& {Morgan}, E.~H. 1996, \procspie, 2808, 59

\bibitem[{{Levine} {et~al.}(1996){Levine}, {Bradt}, {Cui}, {Jernigan},
  {Morgan}, {Remillard}, {Shirey}, \& {Smith}}]{asm96}
{Levine}, A.~M., {Bradt}, H., {Cui}, W., {Jernigan}, J.~G., {Morgan}, E.~H.,
  {Remillard}, R., {Shirey}, R.~E., \& {Smith}, D.~A. 1996, \apjl, 469, L33

\bibitem[{{Manchester} \& {Taylor}(1977)}]{mt77}
{Manchester}, R.~N. \& {Taylor}, J.~H. 1977, Pulsars (San Francisco:
  W.~H.~Freeman)

\bibitem[{{Markwardt} {et~al.}(2004{\natexlab{a}}){Markwardt}, {Galloway},
  {Chakrabarty}, {Morgan}, \& {Strohmayer}}]{mark04b}
{Markwardt}, C.~B., {Galloway}, D.~K., {Chakrabarty}, D., {Morgan}, E.~H., \&
  {Strohmayer}, T.~E. 2004{\natexlab{a}}, The Astronomer's Telegram, 360

\bibitem[{{Markwardt} {et~al.}(2003){Markwardt}, {Smith}, \&
  {Swank}}]{markwardt03}
{Markwardt}, C.~B., {Smith}, E., \& {Swank}, J.~H. 2003, \iaucirc, 8080

\bibitem[{{Markwardt} \& {Swank}(2003)}]{markwardt03a}
{Markwardt}, C.~B. \& {Swank}, J.~H. 2003, \iaucirc, 8144

\bibitem[{{Markwardt} {et~al.}(2004{\natexlab{b}}){Markwardt}, {Swank}, \&
  {Strohmayer}}]{mark04a}
{Markwardt}, C.~B., {Swank}, J.~H., \& {Strohmayer}, T.~E. 2004{\natexlab{b}},
  The Astronomer's Telegram, 353

\bibitem[{{Markwardt} {et~al.}(2002){Markwardt}, {Swank}, {Strohmayer}, {in 't
  Zand}, \& {Marshall}}]{markwardt02}
{Markwardt}, C.~B., {Swank}, J.~H., {Strohmayer}, T.~E., {in 't Zand}, J.
  J.~M., \& {Marshall}, F.~E. 2002, \apjl, 575, L21

\bibitem[{{Narayan} \& {Heyl}(2003)}]{ramesh03}
{Narayan}, R. \& {Heyl}, J.~S. 2003, \apj, 599, 419

\bibitem[{{Nelson} \& {Rappaport}(2003)}]{nr03}
{Nelson}, L.~A. \& {Rappaport}, S. 2003, \apj, 598, 431

\bibitem[{{Nelson} {et~al.}(1986){Nelson}, {Rappaport}, \& {Joss}}]{nrj86}
{Nelson}, L.~A., {Rappaport}, S.~A., \& {Joss}, P.~C. 1986, \apj, 304, 231

\bibitem[{{Pooley}(2004)}]{pooley04}
{Pooley}, G. 2004, The Astronomer's Telegram, 355

\bibitem[{{Remillard}(2004)}]{remillard04}
{Remillard}, R. 2004, The Astronomer's Telegram, 357

\bibitem[{{Roelofs} {et~al.}(2004){Roelofs}, {Jonker}, {Steeghs}, {Torres}, \&
  {Nelemans}}]{roelofs04}
{Roelofs}, G., {Jonker}, P.~G., {Steeghs}, D., {Torres}, M., \& {Nelemans}, G.
  2004, The Astronomer's Telegram, 356

\bibitem[{{Standish} {et~al.}(1992){Standish}, {Newhall}, {Williams}, \&
  {Yeomans}}]{standish92}
{Standish}, E.~M., {Newhall}, X.~X., {Williams}, J.~G., \& {Yeomans}, D.~K.
  1992, in Explanatory Supplement for the Astronomical Almanac, ed. P.~K.
  {Seidelmann} (Mill Valley: University Science), 279

\bibitem[{{Titarchuk}(1994)}]{tit94}
{Titarchuk}, L. 1994, \apj, 434, 570

\bibitem[{{Titarchuk} {et~al.}(2002){Titarchuk}, {Cui}, \& {Wood}}]{tcw02}
{Titarchuk}, L., {Cui}, W., \& {Wood}, K. 2002, \apjl, accepted
  (astro-ph/0207552)

\bibitem[{{Wagoner}(1984)}]{wagoner84}
{Wagoner}, R.~V. 1984, \apj, 278, 345

\bibitem[{{White} \& {Zhang}(1997)}]{wz97}
{White}, N.~E. \& {Zhang}, W. 1997, \apjl, 490, L87

\bibitem[{{Wijnands}(2003)}]{wij03a}
{Wijnands}, R. 2003, in Proceedings of the Symposium "The Restless High-Energy
  Universe", 5-8 May 2003, Amsterdam, The Netherlands, ed. E.~P.~J. {van den
  Heuvel}, J.~J.~M. {in 't Zand}, \& R.~A. M.~J. {Wijers}

\bibitem[{{Wijnands} \& {van der Klis}(1998)}]{wij98b}
{Wijnands}, R. \& {van der Klis}, M. 1998, \nat, 394, 344

\bibitem[{{Wijnands} {et~al.}(2003){Wijnands}, {van der Klis}, {Homan},
  {Chakrabarty}, {Markwardt}, \& {Morgan}}]{wij03}
{Wijnands}, R., {van der Klis}, M., {Homan}, J., {Chakrabarty}, D.,
  {Markwardt}, C.~B., \& {Morgan}, E.~H. 2003, \nat, 424, 44

\bibitem[{{Winkler} {et~al.}(2003){Winkler}, {Courvoisier}, {Di Cocco},
  {Gehrels}, {Gim{\' e}nez}, {Grebenev}, {Hermsen}, {Mas-Hesse}, {Lebrun},
  {Lund}, {Palumbo}, {Paul}, {Roques}, {Schnopper}, {Sch{\" o}nfelder},
  {Sunyaev}, {Teegarden}, {Ubertini}, {Vedrenne}, \& {Dean}}]{integral03}
{Winkler}, C., {et~al.}
  2003, \aap, 411, L1

\end{thebibliography}
\end{document}